\documentclass[a4paper,12pt]{article}
\pdfoutput=1
\usepackage{epsfig}
\usepackage{amssymb}
\usepackage{amsfonts}
\usepackage{amsmath}
\usepackage{euscript}
\usepackage{verbatim}
\usepackage{latexsym}
\usepackage{graphicx}
\usepackage{caption}
\usepackage{float}
\usepackage{subcaption}
\usepackage{wrapfig}
	\usepackage[T1]{fontenc}
	\usepackage{tikz}
	\usetikzlibrary{decorations.pathmorphing}
\usepackage{tikz}
\usetikzlibrary{shapes.geometric, arrows,patterns,snakes}
\tikzstyle{ellip} = [ellipse, minimum width=3cm, minimum height=1cm,text centered, draw=black]

\newskip\humongous \humongous=0pt plus 1000pt minus 1000pt

\newif\ifdtup

\catcode`\@=11

\@addtoreset{equation}{section}

\def\@normalsize{\@setsize\normalsize{15pt}\xiipt\@xiipt
\abovedisplayskip 14pt plus3pt minus3pt%
\belowdisplayskip \abovedisplayskip
\abovedisplayshortskip \z@ plus3pt%
\belowdisplayshortskip 7pt plus3.5pt minus0pt}

\def\small{\@setsize\small{13.6pt}\xipt\@xipt
\abovedisplayskip 13pt plus3pt minus3pt%
\belowdisplayskip \abovedisplayskip
\abovedisplayshortskip \z@ plus3pt%
\belowdisplayshortskip 7pt plus3.5pt minus0pt
\def\@listi{\parsep 4.5pt plus 2pt minus 1pt
     \itemsep \parsep
     \topsep 9pt plus 3pt minus 3pt}}

\relax

\catcode`@=12

\catcode`\@=11

\def\subsection{\@startsection{subsection}{1}{\z@}{3.5ex plus 1ex minus
   .2ex}{2.3ex plus .2ex}{\large\bf}}

\def\SymBoxes#1#2#3#4{\newdimen\un@t \un@t#3%
\raisebox{#1}{\rule{#2\un@t}{#4}\hskip-#2\un@t% lower horizontal
\@tempdimb\un@t \advance\@tempdimb by-#4\@tempcntb#2\relax%
\@whilenum{\@tempcntb>0}\do{%                         % #2 vertical lines
\rule{#4}{\un@t}\hskip\@tempdimb \advance\@tempcntb by\m@ne}%
\hskip-#2\un@t \rule[\un@t]{#2\un@t}{#4}%
\rule[\un@t]{#4}{#4}\hskip-#4%             % upper horizontal line
\rule{#4}{\un@t}}\hskip-#4}                % rightest vertical line
\begin{document}
%\begin{letter}{~}

%%%%%%Define some new commands and  macros
\newcommand{\beq}{\begin{equation}}
\newcommand{\eeq}{\end{equation}}
\newcommand{\bea}{\begin{eqnarray}}
\newcommand{\eea}{\end{eqnarray}}
\newcommand{\beas}{\begin{eqnarray*}}
\newcommand{\eeas}{\end{eqnarray*}}
\newcommand{\defi}{\stackrel{\rm def}{=}}
\newcommand{\non}{\nonumber}
\newcommand{\bquo}{\begin{quote}}
\newcommand{\enqu}{\end{quote}}
%%%%%%%%%%%%%%%%
\renewcommand{\(}{\begin{equation}}
\renewcommand{\)}{\end{equation}}
%%%%%%%%%%%%%%%%%%%%%%%%%%%%%%%%%% definitions
\def \eqn#1#2{\begin{equation}#2\label{#1}\end{equation}}

\def\e{\epsilon}
\def\IZ{{\mathbb Z}}
\def\IR{{\mathbb R}}
\def\IC{{\mathbb C}}
\def\IQ{{\mathbb Q}}
\def\de{\partial}
\def\Tr{ \hbox{\rm Tr}}
\def\H{ \hbox{\rm H}}
\def\HE{ \hbox{$\rm H^{even}$}}
\def\HO{ \hbox{$\rm H^{odd}$}}
\def\K{ \hbox{\rm K}}
\def\Im{ \hbox{\rm Im}}
\def\Ker{ \hbox{\rm Ker}}
\def\const{\hbox {\rm const.}}
\def\o{\over}
\def\im{\hbox{\rm Im}}
\def\re{\hbox{\rm Re}}
\def\bra{\langle}\def\ket{\rangle}
\def\Arg{\hbox {\rm Arg}}
\def\Re{\hbox {\rm Re}}
\def\Im{\hbox {\rm Im}}
\def\exo{\hbox {\rm exp}}
\def\diag{\hbox{\rm diag}}
\def\longvert{{\rule[-2mm]{0.1mm}{7mm}}\,}
\def\a{\alpha}
\def\dag{{}^{\dagger}}
\def\tq{{\widetilde q}}
\def\p{{}^{\prime}}
\def\W{W}
\def\N{{\cal N}}
\def\hsp{,\hspace{.7cm}}

\def\br{\nonumber\\}
\def\IZ{{\mathbb Z}}
\def\IR{{\mathbb R}}
\def\IC{{\mathbb C}}
\def\IQ{{\mathbb Q}}
\def\IP{{\mathbb P}}
\def \eqn#1#2{\begin{equation}#2\label{#1}\end{equation}}

\newcommand{\C}{\ensuremath{\mathbb C}}
\newcommand{\Z}{\ensuremath{\mathbb Z}}
\newcommand{\R}{\ensuremath{\mathbb R}}
\newcommand{\rp}{\ensuremath{\mathbb {RP}}}
\newcommand{\cp}{\ensuremath{\mathbb {CP}}}
\newcommand{\vac}{\ensuremath{|0\rangle}}
\newcommand{\vact}{\ensuremath{|00\rangle}                    }
\newcommand{\oc}{\ensuremath{\overline{c}}}
\newcommand{\bt}{\mathbf{T}}
\newcommand{\bth}{\hat{\mathbf{T}}}

\newcommand{\psiin}{\psi_{0}}
\newcommand{\phiin}{\phi_{1}}
\newcommand{\hin}{h_{0}}
\newcommand{\rh}{r_{h}}
\newcommand{\rb}{r_{b}}
\newcommand{\psibnd}{\psi_{0}^{b}}
\newcommand{\psibndp}{\psi_{1}^{b}}
\newcommand{\phibnd}{\phi_{0}^{b}}
\newcommand{\phibndp}{\phi_{1}^{b}}
\newcommand{\gbnd}{g_{0}^{b}}
\newcommand{\hbnd}{h_{0}^{b}}
\newcommand{\zh}{z_{h}}
\newcommand{\zb}{z_{b}}
\newcommand{\man}{\mathcal{M}}
\newcommand{\ee}{\mathbf{e}}
\newcommand{\real}{\text{Re}}
\newcommand{\imag}{\text{Im}}
\newcommand{\caln}{\mathcal{N}}
\newcommand{\cor}{c_{\mathcal{R}}^{\Omega}}
\newcommand{\boi}{b_{\mathcal{I}}^{\Omega}}
\newcommand{\calg}{\mathcal{G}}
\newcommand{\cald}{\mathcal{D}}
\newcommand{\calo}{\mathcal{O}}
\newcommand{\calh}{\mathcal{H}}
\newcommand{\calr}{\mathcal{R}}
\newcommand{\dslash}{\partial\hspace{-.25cm}/}
\newcommand{\delslash}{\nabla\hspace{-.3cm}/}
\newcommand{\zetabar}{\overline{\zeta}}
\newcommand{\calw}{\mathcal{W}}
\newcommand{\deps}{\delta_{\epsilon}}
\newcommand{\rfal}[1]{\rho^{#1}}
\newcommand{\calor}[1]{\calo\bigl(\rfal{-(#1)}\bigr)}
\newcommand{\khat}{\hat{K}}
\newcommand{\kshat}{\hat{k}}
\newcommand{\ord}[1]{\calo\bigl(r^{-(#1)}\bigr)}
\newcommand{\pihat}{\hat{\mathbf{T}}}
\newcommand{\ud}[2]{_{#1}^{#2}}
\newcommand{\bb}{\mathcal{B}}

\begin{titlepage}
\begin{flushright}
CHEP XXXXX
%ULB-TH/09-10\\
%hep-th/yymmnnn\\
\end{flushright}
\bigskip
\def\thefootnote{\fnsymbol{footnote}}

\begin{center}
{\Large
{\bf Robin Gravity \\ \vspace{0.1in} 
}
}
\end{center}

\bigskip
\begin{center}
{\large  Chethan KRISHNAN$^a$\footnote{\texttt{chethan.krishnan@gmail.com}}, Shubham MAHESHWARI$^a$\footnote{\texttt{shubham.93@gmail.com}}  \\
\vspace{0.1in}
 and P. N. Bala SUBRAMANIAN$^a$\footnote{\texttt{pnbala@cts.iisc.ernet.in}}}
\vspace{0.1in}

\end{center}

\renewcommand{\thefootnote}{\arabic{footnote}}

\begin{center}
%\vspace{0.2cm}

$^a$ {Center for High Energy Physics,\\
Indian Institute of Science, Bangalore 560012, India}\\

\end{center}

\noindent
\begin{center} {\bf Abstract} \end{center}

We write down a Robin boundary term for general relativity. The construction relies on the Neumann result of arXiv:1605.01603 in an essential way. This is unlike in mechanics and (polynomial) field theory, where two formulations of the Robin problem exist: one with Dirichlet as the natural limiting case, and another with Neumann.

%The result is simple, but it uses the Neumann construction of arXiv:1605.01603 in an essential way. This is unlike in mechanics and (polynomial) field theory, where one can get to Robin directly from Dirichlet.

Contribution to the proceedings of the IF-YITP Symposium VI, Phitsanulok, Thailand. 3rd-5th August, 2016.
%, held in Naresuan University, Thailand, August 2016.

\vspace{1.6 cm}
\vfill

\end{titlepage}

\setcounter{footnote}{0}

%%%%%%%%%%%%%%%%%%%%%%%%%%%%%%%%%%%%%%%%%%%%%%%%%%%%%%%%%%%%%%%%%%%%%%%%%%%%%%%%%%%%%%%%%%%%%%
%%%%%%%%%%%%%%%%%%%%%%%%%%%%%%%%%%%%%%%%%%%%%%%%%%%%%%%%%%%%%%%%%%%%%%%%%%%%%%%%%%%%%%%%%%%%%%
\section{Introduction}
The Dirichlet boundary term for general relativity was found by York and Gibbons-Hawking\cite{GHY} long ago, but the Neumann term was only written down recently \cite{Neumann1}. See \cite{Neumann2,Neumann3} for follow-ups.
In this contribution, we will further develop the result of \cite{Neumann1} to construct a well-defined Robin variational problem and construct the general Robin boundary term for general relativity. The Neumann term turns out to be essential for the Robin construction in gravity in a way that it is not, in mechanics and field theory. To clarify this, we outline the various constructions systematically. We also briefly comment on aspects of such a term in asymptotically flat and asymptotically AdS spacetimes. 

This contribution is a small application of the results in \cite{Neumann1, Neumann2, Neumann3}. But the existence of a Robin term for general relativity does not seem to have been appreciated in the literature, so we hope it will be of some use to someone somewhere sometime.

\section{Robin Mechanics}

Let us start by looking at boundary terms in the simplest setting: particle mechanics. Consider the usual Newton action
\bea\label{particleD}
S^{p}_{D}[q] = \int_{T} dt \ L(q,\dot{q}) = \int_{T}dt \ \Biggl(\dfrac{1}{2}\dot{q}^{2} - V(q)\Biggr).
\eea
The superscript $ p $ indicates the action is that of a particle, and the subscript $ D $ denotes that it leads to a well defined variational problem with {\em Dirichlet} boundary condition (in the time direction). To restate the well-known, the variation of the action gives
\bea
\delta S_{D}^{p} = -\int_{T}dt \ \Bigl(\ddot{q} + V'(q)\Bigr)\delta q + (\dot{q} \delta q)\biggr|_{T}.
\eea
If one sets $ q= any \ fixed \ quantity $ at the endpoint\footnote{We keep track of only one boundary, as it is suffices to make our point.} $ T $, the variational problem becomes well posed, and since we are setting $ \delta q|_T =0 $ we call it a Dirichlet problem. 

Note that setting $\dot{q}|_T=0$ in $\delta S^p_D$ is another way to define a valid variational problem, while {\em not} demanding\footnote{Demanding both $\delta q=0=\dot{q}$ fixes both the function and its derivative at the boundary, constraining dynamics uniquely. This is not what we want from a theory: it should allow dynamics, not uniquely fix it.} that $\delta q=0$. We will call this the {\em Special Neumann} boundary condition. 

We would like to find a variational principle where holding $ \delta\dot{q} = 0 $  is well-defined. This is the natural {\em General Neumann} boundary condition, and to accomplish this we add a boundary term to the action:
\bea
S_{N}^{p} &=& S_{D}^{p} - (q\ \dot{q})\biggr|_{T},\\
\Rightarrow \delta S_{N}^{p} &=& -\int_{T}dt \ \Bigl(\ddot{q} + V'(q)\Bigr) \delta q - (q \delta\dot{q})\biggr|_{T}.
\eea
We could restate it in terms of conjugate quantities at the boundary, which leads to a more useful notation later, as
\bea
(q\ \dot{q})\biggr|_{T} \equiv \pi_{T}\, q_{T}, \ \text{where} \ \pi_{T} \equiv \dfrac{\delta S_{D}^{p}}{\delta q_{T}}.
\eea
Note that General Neumann boundary conditions basically mean fixing $\dot{q}=$  {\em any fixed value}, while Special Neumann boundary condition allows only the possibility $ \dot{q}=0$. 

Now, let us consider another boundary term that we could add to $S_D^p$, namely $ S_{1}^{p}= \frac{\xi}{2}q^{2}|_{T} $. This generalizes the Special Neumann boundary condition and leads to what we will call the {\em Special Robin} boundary condition. Upon varying $ S_{D}^{p} +S_{1}^{p} $ we get
\bea
\delta(S_{D}^{p} + S_{1}^{p}) =  -\int_{T}dt \ \Bigl(\ddot{q} + V'(q)\Bigr) \delta q + \Bigl( \dot{q} + \xi \ q\Bigr)\delta q\Bigr|_{T}.
\eea
If we set $\delta q=0$, this is still the Dirichlet variational problem. But we can also set $ (\dot{q} + \xi q)|_{T} = 0 $, which is the Special Robin boundary condition:  holding a linear combination of the position and velocity fixed to zero at the boundary. When $\xi=0$, this reduces to the Special Neumann boundary condition.

What is the Robin analogue of the General Neumann boundary condition? Lets consider adding one more piece to our General Neumann action\footnote{It is possible to set up a General Robin boundary problem for particle mechanics, by starting with the Dirichlet action and never going through the Neumann action. However, this approach does not work for gravity and we find that the Neumann action is crucial for the construction of the Robin action for general relativity. We discuss these matters in Appendix A.}:
\bea
S_{R}^{p} = \int_{T}dt \Biggl(\dfrac{1}{2}\dot{q}^{2} - V(q)\Biggr)  - \Bigl(\dot{q} \, q + \dfrac{\xi}{2}q^{2}\Bigr)\Bigr|_{T},
\eea
which upon varying gives
\bea
\delta S_{R}^{p}  = -\int_{T}dt \ \Bigl(\ddot{q} + V'(q)\Bigr) \delta q - q\,\delta\Bigl(\dot{q} + \xi q\Bigr) \Bigr|_{T}.
\eea 
The variational problem is well defined by setting $ \dot{q} + \xi q = any \ fixed \ value $. This is the {\em General Robin} boundary condition. Again, we could phrase the whole thing as
\bea
\delta S_{R}^{p} = \text{eoms} - q_{T} \,\delta (\pi_{T} + \xi q_{T}) = \text{eoms} - q_{T} \, \delta\Biggl(\dfrac{\delta S_{D}^{p}}{\delta q_{T}} + \xi q_{T}\Biggr).
\eea
The Dirichlet problem can be understood as a variational problem with the position of the particle held arbitrary and fixed at $T$, and the General Neumann problem to be a variational problem with the momentum at $T$ held arbitrary and fixed. The General Robin boundary condition is analogously to be thought of as holding some linear combination of the position and momentum held arbitrary and fixed at $T$. The Dirichlet \cite{GHY} and General Neumann problem \cite{Neumann1} for general relativity are solved, here we would like to fill the gap and formulate the General Robin problem for gravity. There is a bit of a subtlety in this compared to the particle mechanics case (see footnote 3).

But before getting to gravity, we consider the field theory case which is essentially just a fancy rewriting of the particle mechanics case.

\section{Robin Field Theory}

We will start with the action for a scalar field living in a $ D $-dimensional manifold $ (\man,g) $, which again is automatically a Dirichlet action, where we hold the field to be at some fixed value at the boundary
\bea\label{scalarD}
S_{D}[\phi] &=& \int_{\man}d^{D}x \sqrt{-g} \left(-\dfrac{1}{2}g^{\mu\nu}\partial_{\mu} \phi\partial_{\nu}\phi - V(\phi)\right),\\
\Rightarrow \delta S_{D}[\phi] &=& \int_{\man}d^{D}x  \left(\partial_{\mu}(\sqrt{-g} g^{\mu\nu}\partial_{\nu}\phi) - \dfrac{\partial V(\phi)}{\partial\phi}\right)\delta \phi - \int_{\partial\man}d^{D-1}x \sqrt{|\gamma|}n_{\mu} g^{\mu\nu}\partial_{\nu}\phi \,\delta\phi \qquad\ \ \ 
\eea
where $ \gamma $ is the metric on the boundary $ \partial\man $ of $ \man $, and $ n_{\mu} $ is the outward drawn normal to the boundary. The standard procedure, as mentioned earlier is to set the Dirichlet boundary condition $ \delta\phi=0 $, which leads to a well defined variational problem. The Special Neumann case is obtained from the same action while setting the directional derivative $\partial_n\phi \equiv n_{\mu} g^{\mu\nu}\partial_{\nu}\phi=0$ instead. As before, if we work with 
\bea
S_D[\phi]+\int_{\partial\man}d^{D-1}x \sqrt{|\gamma|}\ \dfrac{\xi}{2}\phi^{2}
\eea
it leads to the Special Robin variational problem with $ \partial_{n}\phi + \xi\phi =0$ at the boundary. To get the General Neumann action, we take by direct analogy
\bea
S_{N}[\partial_{n}\phi] &=& S_{D}[\phi] + \int_{\partial\man}d^{D-1}x \sqrt{|\gamma|}(n_{\mu} g^{\mu\nu}\partial_{\nu}\phi)\, \phi\, \\
\Rightarrow \delta S_{N}[\partial_{n}\phi] %&=& \text{eom} + \int_{\partial\man}d^{D-1}x\,  \sqrt{|\gamma|}\,\delta(n_{\mu} g^{\mu\nu}\partial_{\nu}\phi )\phi \nonumber\\
&=& \text{eom} + \int_{\partial\man}d^{D-1}x \sqrt{|\gamma|}\; \delta(\partial_{n}\phi) \,\phi.
\eea
The variational problem here is well defined by holding $ \delta(\partial_{n}\phi) = 0 $. The scalar field theory can also be well posed as a General Robin boundary problem:
\bea
S_{R}[\phi] = S_{D}[\phi] + \int_{\partial\man}d^{D-1}x \sqrt{|\gamma|} \left(\phi\, \partial_{n }\phi + \dfrac{\xi}{2}\phi^{2}\right)=S_{N}[\phi]+\int_{\partial\man}d^{D-1}x \sqrt{|\gamma|}\ \dfrac{\xi}{2}\phi^{2},
\eea
which will lead to holding $ \partial_{n}\phi + \xi\phi = any\ fixed\ value $ at $ \partial\man $. We worked with the scalar for simplicity, but this generalizes trivially to the gauge field as well.

\section{Robin Gravity}
We can now proceed to look for a boundary term that gives a consistent Robin boundary problem for gravity. Let us, as usual, start with the Einstein-Hilbert action on a $ D $-dimensional manifold $ (\man,g) $ along with the Gibbons-Hawking York boundary term, which leads to Dirichlet gravity
\bea
S_{D} = S_{EH}+S_{GHY} = \dfrac{1}{2\kappa} \int_{\man}d^{D}x \, \sqrt{-g}(R-2\Lambda)+ \dfrac{1}{\kappa} \int_{\partial\man} d^{D-1}y \sqrt{|\gamma|} \, \epsilon \, \Theta,
\eea
where $ \kappa = 8\pi G_{N} $, $ R $ is the Ricci scalar and $ \Lambda $ is a cosmological constant. Also, $ \gamma_{ij} = g_{\mu\nu}e^{\mu}_{i}e^{\nu}_{j} $ is the induced metric on the boundary $ \partial\man $ and $ e^{\mu}_{i}=\frac{\partial x^{\mu}}{\partial y^{i}} $ projects the bulk coordinates $ x^{\mu} $ to the boundary coordinates $ y^{i} $. The extrinsic curvature of the boundary is given by
\bea
\Theta_{ij} =  \dfrac{1}{2}(\nabla_{\mu}n_{\nu} + \nabla_{\nu}n_{\mu})e^{\mu}_{i}e^{\nu}_{j},
\eea
where $ n_{\mu} $ is the outward drawn unit normal to the boundary, and $ \epsilon= \pm 1 $ distinguishes the boundary between time-like and space-like boundaries respectively.

The variation of Dirichlet action yields
%\bea
%\delta S_{D} &=& \dfrac{1}{2\kappa} \int_{\man}d^{D}x \, \sqrt{-g}\left(R_{\mu\nu}-\dfrac{1}{2}g_{\mu\nu}+\Lambda g_{\mu\nu}\right) \delta g^{\mu\nu} \nonumber\\
%&& \qquad\qquad\qquad\qquad  - \dfrac{1}{\kappa} \int_{\partial\man} d^{D-1}y \sqrt{|\gamma|} \epsilon\left(\delta\Theta +\dfrac{1}{2}\Theta^{ij}\delta\gamma_{ij}\right),
%\eea
%where $ \gamma_{ij} = g_{\mu\nu}e^{\mu}_{i}e^{\nu}_{j} $ is the induced metric on the boundary $ \partial\man $ and $ e^{\mu}_{i}=\frac{\partial x^{\mu}}{\partial y^{i}} $ is the coordinate transformation relating the boundary coordinates $ y^{i} $ to the bulk coordinates $ x^{\mu} $. The extrinsic curvature of the boundary is given by
%\bea
%\Theta_{ij} =  \dfrac{1}{2}(\nabla_{\mu}n_{\nu} + \nabla_{\nu}n_{\mu})e^{\mu}_{i}e^{\nu}_{j},
%\eea
%where $ n_{\mu} $ is the outward drawn unit normal to the boundary, and $ \epsilon= \pm 1 $ distinguishes the boundary between time-like and space-like boundaries respectively. For a well defined variational principle in the standard Dirichlet sense of holding $ \gamma_{ij} $ fixed, the $ \delta\Theta $ term has to be taken care of with a boundary term, \'a la Gibbons-Hawking-York
%\bea
%S_{GHY} = \dfrac{1}{\kappa} \int_{\partial\man} d^{D-1}y \sqrt{|\gamma|} \epsilon \Theta,
%\eea
%on variation which yields
%\bea
%\delta S_{GHY} = \dfrac{1}{\kappa} \int_{\partial\man} d^{D-1}y \sqrt{|\gamma|} \epsilon \left(\delta\Theta+\dfrac{1}{2}\Theta\gamma^{ij}\delta\gamma_{ij}\right) .
%\eea
%The total Dirichlet gravitational action is given by $ S_{D} = S_{EH} + S_{GHY} $, the variation of which is
\bea
\delta S_{D} &=& \delta S_{EH} + \delta S_{GHY}\nonumber \\
&=&  \dfrac{1}{2\kappa} \int_{\man}d^{D}x \, \sqrt{-g}\left(G_{\mu\nu}+\Lambda g_{\mu\nu}\right) \delta g^{\mu\nu} - \dfrac{1}{2\kappa} \int_{\partial\man} d^{D-1}y \sqrt{|\gamma|} \epsilon\left( \Theta^{ij} -\Theta \gamma^{ij}\right)\delta\gamma_{ij},\qquad
\eea
where $ G_{\mu\nu } = R_{\mu\nu} -\frac{1}{2}g_{\mu\nu}R $ is the Einstein tensor. The variational problem is well defined with the boundary metric held fixed, and we can think of $ S_{D} = S_{D}[\gamma_{ij}] $ as a functional of the boundary metric.

We can define a canonical conjugate of the boundary metric as
\bea
\pi^{ij} \equiv \dfrac{\delta S_{D}}{\delta \gamma_{ij}} = - \dfrac{1}{2\kappa}\sqrt{|\gamma|} \epsilon\left( \Theta^{ij} -\Theta \gamma^{ij}\right),
\eea
using which we can rewrite the variation of $ S_{D} $ in a simpler form
\bea
\delta S_{D} =  \dfrac{1}{2\kappa} \int_{\man}d^{D}x \, \sqrt{-g}\left(G_{\mu\nu}+\Lambda g_{\mu\nu}\right) \delta g^{\mu\nu} + \int_{\partial\man} d^{D-1}y \,\pi^{ij} \,\delta\gamma_{ij}.
\eea

We also note that holding $ \pi^{ij} = 0 $ here leads to the Special Neumann boundary condition for gravity. This is sometimes described as the Neumann problem for gravity in the literature, even though it is a special case of the general situation.

As was discussed in \cite{Neumann1,Neumann2,Neumann3}, an action which is well defined in terms of General Neumann boundary condition can be defined as 
\bea
S_{N} &=& S_{EH} + S_{GHY} - \int_{\partial\man} d^{D-1}y \,\pi^{ij} \,\gamma_{ij}\\
&=&  \dfrac{1}{2\kappa} \int_{\man}d^{D}x \, \sqrt{-g}(R-2\Lambda) +  \dfrac{4-D}{2\kappa} \int_{\partial\man} d^{D-1}y \sqrt{|\gamma|} \epsilon \Theta,
\eea
the variation of which is given by
\bea
\delta S_{N} =  \dfrac{1}{2\kappa} \int_{\man}d^{D}x \, \sqrt{-g}\left(G_{\mu\nu}+\Lambda g_{\mu\nu}\right) \delta g^{\mu\nu} + \int_{\partial\man} d^{D-1}y \,\gamma_{ij}\,\delta\pi^{ij}.
\eea
Here, instead of holding the boundary metric fixed, the quantity $ \pi^{ij} $ is held fixed, letting the boundary metric fluctuate. The quantity $ \pi^{ij} $ is termed boundary stress tensor density (also, sometimes as quasi-local stress tensor density\cite{BrownYork}). The Neumann boundary condition can be thought of as looking at solutions holding the boundary stress tensor density fixed, i.e. $ S_{N} = S_{N}[\pi^{ij}] $. 

Now we turn to Special Robin. Adding a boundary term $ S_{b} = 2\zeta \int d^{D-1}y \sqrt{|\gamma|} $ to the Dirichlet action, it is straightforward to again check that we will have a  variational problem well defined under the Special Robin boundary condition, $ \pi^{ij}+\zeta \sqrt{|\gamma|} \gamma^{ij} = 0 $. 

In order to have the action be a well defined variational problem under General Robin boundary condition, we need to add a boundary term which will ensure that $ \pi^{ij} + \xi \sqrt{|\gamma|}\gamma^{ij} $ held arbitrary and fixed\footnote{The explicit presence of $ \sqrt{|\gamma|} $ is not of much worry, as one can see, $ \pi^{ij} $ is also defined implicitly with the same factor.} leads to a consistent variational problem. To get such an action, we go through the Neumann action like we did in the mechanics and field theory cases. Note that unlike in those cases, in gravity we cannot get to Robin from Dirichlet bypassing Neumann\footnote{See discussion in the Appendix for some elaboration on this.}. In other words, going through Neumann is not an option but a necessity  in the case of gravity. 

In any event, the result is
\bea
\label{robin}
S_{R} &=&  \dfrac{1}{2\kappa} \int_{\man}d^{D}x \, \sqrt{-g}(R-2\Lambda) \nonumber\\
&& \hspace{2.5cm} +  \dfrac{4-D}{2\kappa} \int_{\partial\man} d^{D-1}y \sqrt{|\gamma|} \epsilon \Theta - \xi(D-3)\int_{\partial\man} d^{D-1}y\sqrt{|\gamma|}.\ \qquad \
\eea
Varying the action, and using the key relation
\bea
(D-3) \sqrt{|\gamma|}\gamma^{ij}\delta\gamma_{ij} = 2\,\delta(\sqrt{|\gamma|} \gamma^{ij})\,\gamma_{ij}, \label{key}
\eea
we can show that
\bea
\delta S_{R} =  \dfrac{1}{2\kappa} \int_{\man}d^{D}x \, \sqrt{-g}\left(G_{\mu\nu}+\Lambda g_{\mu\nu}\right) \delta g^{\mu\nu} - \int_{\partial\man} d^{D-1}y \,\delta(\pi^{ij}+\xi \sqrt{|\gamma|} \gamma^{ij})\,\gamma_{ij}.
\eea
The action \eqref{robin} is what we call the Robin action for gravity.

\section{Comments}

{\bf Hamiltonian Formulation}

\noindent
We will now write down the Robin Gravity action in the Hamiltonian formulation. Using the fact that the action in \eqref{robin} is the same as that of Neumann gravity, except for an additional boundary cosmological constant term, we can directly write down the action in terms of canonical variables for Robin gravity \cite{Neumann2}:
\bea
S_{R} &=& \int_{\man}d^{D}x\left(p^{ab}\dot{h}_{ab}- N H -N_{a}H^{a}\right) \nonumber\\
&& \qquad\qquad\qquad + \int_{\bb}d^{D-1}y \sqrt{\sigma}\left(N\left(\dfrac{\varepsilon}{2}-\xi(d-3)\right) - N^{a}j_{a}+ \dfrac{N}{2}s^{ab}\sigma_{ab}\right).
\eea
We will not elaborate on the (completely standard) notations here, they can be found in, eg. \cite{Neumann2}. 
%where $ \sqrt{\sigma}\varepsilon,\ \sqrt{\sigma}j_{a} $ and $ N\sqrt{\sigma} s^{ab}/2 $ are the canonical conjugates to $ N,\ N^{a} $ and $ \sigma_{ab} $, the exact expressions of which are given by
%\bea \varepsilon &=& \dfrac{k}{\kappa}, \ \ j_{a} = \dfrac{2}{\sqrt{h}}r_{b}p^{b}_{\; a}\\ s^{ab} &=& \dfrac{1}{\kappa} \left(k^{ab} - \left(\dfrac{r^{c}\partial_{c} N}{N} +k \right)\sigma^{ab}\right). \eea
% Also, $ p^{ab} $ is the canonical conjugate to $ h_{ab} $, $ H $ and $ H^{a} $ are the Hamiltonian and momentum constraints, $ k_{ab} = e^{\alpha}_{a}e^{\beta}_{b}\nabla_{\alpha}r_{\beta} $, $ e^{\alpha}_{a}=\frac{\partial x^{\alpha}}{\partial y^{a}} $ being the projector from coordinates on $ \man $ to $ \Sigma_{t} $, and $ r_{\alpha} $ is the radially outward drawn unit normal.

\noindent
{\bf Asymptotically Flat Space-times}

\noindent
If one goes about naively computing the classical action for any asymptotically flat space-time(AFS), its bound to run into divergences. The usual procedure to deal with in AFS is to do a background subtraction, which involves holding the induced metric at the boundary the same for the background and the datum. For \eqref{robin}, this means that the boundary cosmological constant term drops off and we end up with the same result as what one would get from a pure Neumann boundary term, see \cite{Neumann2}. This means that various discussions there on thermodynamics, horizons, etc \cite{Krishnan} also immediately apply to the background subtracted case here.

%\bea S_{R} = \dfrac{1}{2\kappa}\int_{\man}d^{D}x\sqrt{-g}\, R + \dfrac{4-D}{2\kappa}\int_{\bb} d^{D-1}y\, \sqrt{-\gamma} (\Theta -\Theta_{0}), \eea
%where $ \Theta_{0} $ is the extrinsic curvature of $ \bb $ embedded in flat space (we will be looking at only time-like boundaries for this analysis). The extra piece that was present in \eqref{robin} is subtracted out. Thus we recover a regulated action which in fact is the same as that in (5.22) of \cite{Neumann2}, whereby, we can directly write down Robin gravity action in ADM variables as
%\bea S_{R} &=& \int_{\man}d^{D}x\left(p^{ab}\dot{h}_{ab}- N H -N_{a}H^{a}\right)+ \int_{\mathcal{H}}d^{D-1}x \sqrt{\sigma} \left(\dfrac{r^{a}\partial_{a}N}{\kappa} + \dfrac{2 N^{a}r^{a} p_{ab}}{\sqrt{h}}\right) \nonumber\\ && \qquad\qquad + \int_{\bb}d^{D-1}y \sqrt{\sigma}\left(\dfrac{N}{2}(\varepsilon - \varepsilon_{0}) - N^{a}(j_{a}-j_{a\, 0})+ \dfrac{N}{2}(s^{ab}-s^{ab}_{0})\sigma_{ab}\right).\eea
%We have added a horizon piece as in \cite{Neumann2}. Given that the background subtracted action in the canonical and covariant form for Robin gravity is the same as that of Neumann gravity, it is evident that the Smarr formula and the First Law of black hole thermodynamics holds in the case of Robin gravity also, and we will not redo the entire exercise here.

\noindent
{\bf AdS}

\noindent
We will now look at asymptotically  AdS$_{d+1} $ spaces\footnote{We set $ D=d+1 $ for convenience.}. We follow the notations of \cite{Neumann3}. 
%, given by
%\bea
%ds^{2} &=& \mathcal{G}_{\mu\nu}dx^{\mu}dx^{\nu} = \dfrac{l^{2}}{z^{2}}(dz^{2}+g_{ij} (x,z)dx^{i}dx^{j}),\\
%\text{where}\ \ g(x,z) &=& g_{0} + z^{2}g_{2} +\dots+ z^{d}g_{d} + z^{d}h_{d} \log z^{2} +\mathcal{O}(z^{d+1}).
%\eea
%The cosmological constant is related to the AdS radius though the relation $ \Lambda = -\frac{d(d-1)}{2l^{2}} $, and we will set $ l=1 $ in all the computations that follow. We can do a coordinate transformation $ \rho= z^{2} $ as only even powers appear in the expansion
%\bea
%ds^{2} &=& \dfrac{1}{4\rho^{2}}d\rho^{2} + \dfrac{1}{\rho}g_{ij}(x,\rho)dx^{i}dx^{j}, \\
%g(x,\rho) &=& g_{0} +\rho g_{2}+ \dots +\rho^{\frac{d}{2}}g_{d} +\rho^{\frac{d}{2}} h_{d} \log \rho 
%\eea
%From the Neumann boundary condition analysis of AdS in \cite{Neumann3}, it can be seen that in the case of asymptotically AdS spaces, we could write the renormalized Neumann action as a boundary Legendre transform of the renormalized Dirichlet action, which is obtained by the means of holographic renormalization (see \cite{Skenderis} for a detailed review on this topic). In essence, 

The renormalized Neumann action is given by
\bea
S_{N}^{ren} = S_D^{ren} - \int_{\partial\man} d^{d}x\, \pi^{ij}_{ren} \gamma_{ij},
\eea
where $ \pi^{ij}_{ren} $ is the renormaized boundary stress tensor density.
% The term in the action $ S_{ct}^{(d)} $ is evaluated specifically for each $ d $, to give the renormalized boundary stress tensor density.

The boundary stress tensor is related to $ \pi^{ij} $ as\footnote{The boundary we are looking at is time-like $ \epsilon = +1 $, in the entire AdS discussion.}
\bea
T^{ij} = -\dfrac{2}{\sqrt{-\gamma}}\pi^{ij} = \dfrac{1}{\kappa}(\Theta^{ij} -\Theta\, \gamma^{ij}) ,
\eea
and the renormalized stress tensor $ T^{ij}_{ren} $ is obtained from $ \pi^{ij}_{ren} $ in the same way. The variation of renormalized Neumann action gives
\bea
\delta S_{N}^{ren} &=& \text{eq. of motion }\; - \int_{\partial\man} d^{d}x \, \gamma_{ij} \delta \pi^{^{ij}}_{ren} \nonumber\\
&=&  \text{eq. of motion }\; + \dfrac{1}{2} \int_{\partial\man} d^{d}x \,g_{0\;ij}\delta(\sqrt{-g_{0}} \mathcal{T}^{ij}),
\eea
where $ \mathcal{T}^{ij} $ is the true renormalized stress tensor (of the boundary CFT) and is given by\cite{Holren}
\bea
\mathcal{T}_{ij} &=& \lim_{\epsilon\rightarrow 0}\left(\dfrac{1}{\epsilon^{\frac{d}{2}-1}} T_{ij}^{ren}[\gamma]\right) = \lim_{\epsilon\rightarrow 0}\left(-\dfrac{2}{\sqrt{g(x,\epsilon)}}\dfrac{\delta S_{D}^{ren}}{\delta g^{ij}}\right)= - \dfrac{2}{\sqrt{-g_{0}}}\dfrac{\delta S_{D}^{ren}}{\delta g^{ij}_{0}}.
\eea

We can write down the Robin gravity action specific to AdS as
\bea
S_{R}^{ren} = S_D^{ren} - \int_{\partial\man} d^{d}x\, \pi^{ij}_{ren} \gamma_{ij} +  \dfrac{(d-2)}{2}\xi\int_{\partial\man}d^{d}x \sqrt{-g_{0}} ,
\eea
which upon variation gives
\bea
\delta S_{R} = \text{eq. of motion} + \dfrac{1}{2} \int_{\partial\man} d^{d}x \,g_{0\,ij}\ \delta\biggl(\sqrt{-g_{0}} (\mathcal{T}^{ij} + \xi g_{0}^{ij}) \biggr), 
\eea
with the variational principle well defined by holding $ \sqrt{-g_{0}} (\mathcal{T}^{ij} + \xi g_{0}^{ij}) = any\ fixed\ quantity$. 

The essential difference between flat space and AdS is that here the variational principle is best formulated in terms quantities that are intrinsic to the field theory: in other words, in terms of a combination of the $g_0$ and the $g_d$ in the Fefferman-Graham expansion (see \cite{Neumann3, Holren}) instead of induced metric $\gamma$. Note that $\mathcal{T}^{ij}$ is determined in terms of them \cite{Holren}.

%{\bf The reason for using $ \sqrt{-g_{0}} $ instead of $ \sqrt{-\gamma} $ has to be remarked upon. Also, except for the last term, the action is the same as Neumann, and may be have to remark on on-shell actions for known solutions. } 

\section*{Acknowledgments}

We thank the participants and organizers of the IF-YITP Symposium VI, especially Burin Gumjudpai, Matthew Lake and Shingo Takeuchi for an enjoyable conference.

\appendix

\section{Another Path to Robin?}

In the particle mechanics and field theory cases, there exists a direct path from the Dirichlet action to the General Robin action. Let us start with the particle mechanics problem. To the Dirichlet action in \eqref{particleD}, add a boundary term $ S_{B}^{p} = \frac{\xi}{2} \dot{q}^{2}\bigr|^{T} $, the variation of the sum of two gives
\bea
\delta(S_{D}^{p}+S_{B}^{p}) =  -\int_{T}dt \ \Bigl(\ddot{q}^{2} + V'(q)\Bigr) \delta q + \dot{q}\,\delta\Bigl(\xi\dot{q} +  q\Bigr) \Bigr|_{T}.
\eea
This is clearly a well-defined General Robin variational principle. This sort of thing extends trivially to the field theory case as well. Simply consider the following addition to the Dirichlet field theory action:
\bea
S_{R}' &=& \int_{\man}d^{D}x \sqrt{-g} \left(-\dfrac{1}{2}g^{\mu\nu}\partial_{\mu} \phi\partial_{\nu}\phi - V(\phi)\right) - \dfrac{\xi}{2}\int _{\partial\man}d^{D-1}x\sqrt{|\gamma|} (\partial_{n}\phi)^{2},\\
\Rightarrow \delta S_{R}' &=&  \int_{\man}d^{D}x  \left(\partial_{\mu}(\sqrt{-g} g^{\mu\nu}\partial_{\nu}\phi) - \dfrac{\partial V(\phi)}{\partial\phi}\right)\delta \phi \nonumber +\\
& &\qquad- \int_{\partial\man}d^{D-1}x \sqrt{|\gamma|} \partial_{n}\phi \;\delta\left(\phi + \xi \partial_{n}\phi\right)
\eea

For the case of gravity, one might think that an analogous boundary piece can be added to  the Dirichlet (Gibbons-Hawking-York) action to produce the General Robin action. This however does not seem to work: for gravity, we find that going through the Neumann action seems to be essential to obtain the Robin action. We describe why this is so, below. 

There are two possible terms that could be constructed out of $ \pi_{ij} $'s that are quadratic, namely, $ (\pi^{ij}\gamma_{ij})^{2} $ and $ \pi^{ij}\gamma_{jk}\pi^{kl}\gamma_{li}$. Also, one has to remember that $ \pi^{ij} $ internally contains a $ \sqrt{|\gamma|} $ factor, so it would be more advisable to write boundary terms using $ T^{ij} $, the boundary stress tensor\footnote{In the following we use the defintions $ [\Tr(T)]^{2}=(T^{ij}\gamma_{ij})^{2} $ and $ \Tr(T^{2}) = T^{ij}\gamma_{jk}T^{kl}\gamma_{li} $.}. Let us look at the variation of the first candidate, modulo the constants
\bea
\delta S_{1} &=& \int_{\partial\man}d^{D-1}x \;\delta\bigl[\sqrt{|\gamma|} \;[\Tr(T )]^{2}\bigr] \nonumber\\
&=&  \int_{\partial\man}d^{D-1}x \sqrt{|\gamma|}\left(\dfrac{1}{2}[\Tr(T)]^{2} \gamma^{ij}\delta\gamma_{ij} + 2 \,\Tr(T)\bigl( T^{ij}\delta\gamma_{ij} + \gamma_{ij} \delta T^{ij}\bigr)\right).
\eea
This can be written in terms of $ \pi^{ij} $'s as 
\bea
\delta S_{1} &=& 4\int_{\partial\man}d^{D-1}x \;\delta \left[\dfrac{1}{\sqrt{|\gamma|}} (\pi^{ij}\gamma_{ij})^{2}\right]\nonumber\\
&=& 4\int_{\partial\man}d^{D-1}x \;\dfrac{1}{\sqrt{|\gamma|}} \left(-\dfrac{1}{2}(\pi^{ij}\gamma_{ij})^{2} \gamma^{kl}\delta\gamma_{kl} + 2(\pi^{ij}\gamma_{ij}) \pi^{kl}\delta\gamma_{kl} + 2(\pi^{ij}\gamma_{ij}) \gamma_{kl}\delta\pi^{kl}\right).\qquad
\eea
The variation of the second candidate term gives,
\bea
\delta S_{2} = &=& \int_{\partial\man}d^{D-1}x \;\delta\bigl[\sqrt{|\gamma|} \;\Tr(T ^{2})\bigr] \nonumber\\
&=&  \int_{\partial\man}d^{D-1}x \sqrt{|\gamma|}\left(\dfrac{1}{2}\Tr(T^{2}) \gamma^{ij}\delta\gamma_{ij} +2\, T^{ij} \delta T_{ij} + 2\, T^{ij} T_{j}^{\; k} \delta\gamma_{ik}\right).
\eea
This can be written in terms of $ \pi^{ij} $'s as
\bea
\delta S_{2} &=& 4\int_{\partial\man}d^{D-1}x \;\delta\left[\dfrac{1}{\sqrt{|\gamma|}} \;(\pi^{ij}\gamma_{jk}\pi^{kl}\gamma_{li})\right] \nonumber\\
&=& 4\int_{\partial\man}d^{D-1}x \;\dfrac{1}{\sqrt{|\gamma|}} \Bigl(-\dfrac{1}{2}(\pi^{ij}\gamma_{jm}\pi^{mn}\gamma_{ni}) \gamma^{kl}\delta\gamma_{kl} \nonumber \\
& & \hspace*{4cm} + 2\pi^{ij}\gamma_{jk} \pi^{kl}\delta\gamma_{li} + 2\pi^{ij}\gamma_{jk} \gamma_{li}\delta\pi^{kl}\Bigr).\qquad
\eea
To allow the most general possibility, let us consider adding a combination of these two candidate terms with arbitrary coefficients to the Dirichlet action:
\bea
S'_{R} = S_{D} +\xi  \int_{\partial\man}d^{D-1}x \;\dfrac{1}{\sqrt{|\gamma|}} \;(\pi^{ij}\gamma_{ij})^{2} + \zeta \int_{\partial\man}d^{D-1}x \;\dfrac{1}{\sqrt{|\gamma|}} \; \pi^{ij}\gamma_{jk}\pi^{kl}\gamma_{li}.
\eea
Upon variation this yields
\bea
\delta S'_{R} &=& \text{eom} + \int_{\partial\man}d^{D-1}x \;\pi^{ij}\Biggl\{\Bigl[\delta^{k}_{i}\delta^{l}_{j} +\dfrac{\xi}{\sqrt{|\gamma|}}\Bigl(-\dfrac{1}{2}\pi^{mn} \gamma_{mn}\gamma_{ij}\gamma^{kl}+ 2 \pi^{mn}\gamma_{mn} \delta^{k}_{i}\delta^{l}_{j} \Bigr) \nonumber\\
& &\hspace{5cm} + \dfrac{\zeta}{\sqrt{|\gamma|}}\Bigl(-\dfrac{1}{2} \pi^{mn}\gamma_{im}\gamma_{jn}\gamma^{kl}+2 \pi^{nl}\gamma_{jn}\delta^{k}_{i} \Bigr) \Bigr]\delta \gamma_{kl} \nonumber\\
& & \hspace{5cm }+ \dfrac{2}{\sqrt{|\gamma|}}\Bigl[\xi \gamma_{ij} \gamma_{kl} + \zeta \gamma_{ik}\gamma_{jl}\Bigr] \delta\pi^{kl} \Biggr\}
\eea
For this to reduce to a General Robin variation, we need the coefficient of  $\delta\pi^{kl}$  to be some number times the coefficient of $\delta \gamma_{kl}$. This is clearly impossible for any choice of $\zeta $ and $ \xi $.

The essential difference between mechanics/field theory and gravity is that here, the $\sqrt{|\gamma|}$ term shows up, which is essentially non-polynomial. This makes the Neumann term an essential intermediate step in our path to Robin: there does not seem to be direct path to it from the Dirichlet (Gibbons-Hawking) boundary term. 

Another way to state the same observation is that one can view both Dirichlet and Neumann boundary conditions as limits of Robin in mechanics and field theory, but in general relativity only the Neumann boundary condition can be viewed as a limit of Robin. At the technical level, the problem is that for $\gamma$, the key relation \eqref{key} holds, but for $\pi$ there is no such relation.

\end{document}